\documentclass[conference]{IEEEtran}
\ifCLASSINFOpdf
\else
\fi

\usepackage{color}
\usepackage{amsmath,amssymb,amscd}
\usepackage{subfigure, epsfig}
\usepackage{pstricks}
\usepackage{pstricks-add}
\usepackage{pst-plot}

\newtheorem{theorem}{Theorem}
\newtheorem{definition}[theorem]{Definition}

\newcommand{\ul}{\underline}

\newcommand{\PPP}{\mathbb{P}}
\newcommand{\E}{\mathbb{E}}

\newcommand{\G}{\mathcal{G}}

\newcommand{\SSS}{\mathcal{S}}

\newcommand{\T}{\mathcal{T}}

\newcommand{\PP}{\mathcal{P}}

\newcommand{\mc}{\mathcal}

\begin{document}

\title{Implicit cooperation in distributed energy-efficient networks}

\author{\IEEEauthorblockN{Ma\"{e}l Le Treust\IEEEauthorrefmark{1},
Samson Lasaulce\IEEEauthorrefmark{1} and
M\'{e}rouane Debbah \IEEEauthorrefmark{3}}
\IEEEauthorblockA{\IEEEauthorrefmark{1}Laboratoire des Signaux et Syst\`{e}mes,
CNRS  - Universit\'{e} Paris-Sud 11 - Sup\'{e}lec,
91191, Gif-sur-Yvette Cedex, France\\
Email: \{letreust\},\{lasaulce\}@lss.supelec.fr}
\IEEEauthorblockA{\IEEEauthorrefmark{3}
Chaire Alcatel-Lucent,
SUPELEC,
91190 Gif-sur-Yvette, France\\
Email: debbah@supelec.fr}}

%\author{M. Le Treust, S. Lasaulce, and M. Debbah \thanks{M. Le Treust, S. Lasaulce are with LSS-CNRS-Supelec-Universit\'e de Paris 11 ; 3, rue Joliot-Curie, 91190 Gif-sur-Yvette, France and M. Debbah is with Chaire Alcatel-Supelec ; 3, rue Joliot-Curie, 91190 Gif-sur-Yvette, France}}

% \thanks{M. Le Treust, S. Lasaulce are with LSS-CNRS-Supelec-Universit\'e de Paris 11 ; 3, rue Joliot-Curie, 91190 Gif-sur-Yvette, France and M. Debbah is with Chaire Alcatel-Supelec ; 3, rue Joliot-Curie, 91190 Gif-sur-Yvette, France}
% Les affiliations (par ordre croissant des numéros d'affiliation) séparées par \and

%\institute{LSS-CNRS-Supelec-Universit\'e de Paris 11 ; 3, rue Joliot-Curie, 91190 Gif-sur-Yvette, France \\
%\email{ \{mael.letreust,lasaulce\}@lss.supelec.fr}

\maketitle

%\tableofcontents

\begin{abstract}
We consider the problem of cooperation in distributed wireless
networks of selfish and free transmitters aiming at maximizing their
energy-efficiency. The strategy of each transmitter consists in
choosing his power control (PC) policy. Two scenarios are
considered: the case where transmitters can update their power
levels within time intervals less than the channel coherence time
(fast PC) and the case where it is updated only once per time
interval (slow PC). One of our objectives is to show how cooperation
can be stimulated without assuming cooperation links between the
transmitters but only by repeating the corresponding PC game and by
signals from the receiver. In order to design efficient PC policies,
standard and stochastic repeated games are respectively exploited to
analyze the fast and slow PC problems. In the first case a
cooperation plan between transmitters, that is both efficient and
relies on mild information assumptions, is proposed. In the second
case, the region of equilibrium utilities is derived from very
recent and powerful results in game theory.
\end{abstract}
%
%
%\begin{IEEEkeywords}
%Cognitive radio, energy-efficiency, Folk theorem, Nash equilibrium,
%power control games, repeated games, subgame perfect equilibrium.
%
%\end{IEEEkeywords}

%------------------------------------------------------------
%------------------------------------------------------------
%------------------------------------------------------------
%------------------------------------------------------------

\section{Introduction}
\label{sec:intro}

In the wireless literature, when it is referred to cooperative
networks, this generally means that some nodes in the network act as
relays in order to help other nodes (the sources or transmitters) to
better communicate with their respective destination or receiver
nodes. This idea has been formalized in information theory in
\cite{vandermeulen-aap-1971}\cite{cover-it-1979} for the relay channel, for the
cooperative multiple access channel (MAC) \cite{willems-it-1983}, and for
other types of cooperative channels during the last decade
(\cite{DaboraServetto06}\cite{kramer-it-2005} etc). The vast majority of these papers address
centralized networks and cooperation between nodes is based on the
existence of physical links between some nodes. In the present paper
we consider the case of decentralized or distributed networks with
implicit cooperation. By decentralized/distributed, we mean that the
nodes are assumed to be free decision makers who decide by
themselves what is good for them and can ignore possible
recommandations from central nodes (namely the power control policy
in our case). By implicit, we mean that nodes cooperate without
using dedicated cooperation channels between nodes. To be more
concrete, we consider multiple access channels where no link between
the transmitters is assumed and transmitters are modeled by selfish
players aiming at maximizing the energy-efficiency of their
communication. A very simple and pragmatic way of knowing to what
extent a communication is energy-efficient has been proposed by
\cite{shah-pimrc-1998}\cite{goodman-pc-2000}. The authors of these
articles define energy-efficiency as the net number of information
bits that are transmitted without error per time unit (goodput) to
the transmit power level. More specifically, the authors analyze the
problem of distributed power control (PC) in flat fading multiple
access channels. The problem is formulated as a non-cooperative
one-shot game where the players are the transmitters, the strategy
of a given player is his transmit power for a given channel
realization, and his payoff/reward/utility function is
energy-efficiency of his communication with the receiver.
Unfortunately, the Nash equilibrium (NE) resulting from this game is
generally inefficient.

In the papers on energy-efficient power control cited above and
related papers (e.g., \cite{meshkati-jsac-2006}\cite{lasaulce-twc-2009}), the used game-theoretic
framework is the one of static or one-shot games for which
transmitters are assumed to interact once per block and from block
to block in an independent manner; the block duration is assumed to
be less than the channel coherence time. In practice, there will be
some scenarios where transmitters can update their power level
several times within a block or/and are active over several and
possibly many blocks. In game theory, it is well known that this
feature can change the behavior of the players and incite them to
cooperate while staying selfish \cite{Aum81}. The corresponding
game-theoretic framework is then the one of dynamic games. In this
paper we propose to model the distributed energy-efficient power
control problem by exploiting two types of repeated games (RG),
namely the standard and stochastic RG, which are special cases of
dynamic games. In standard RG \cite{Sorin92}, the same game is repeated a
certain number of times. In stochastic RG \cite{Shapley53}, players' utilities depend on
 a certain state (or parameters) which vary over time according to a stochastic process. We use
standard RG to analyze scenarios where transmitters can update their
powers several times within each block (the same game is therefore
repeated within a block) and stochastic RG for scenarios where
transmitters update their powers once per block (the game is
therefore parameterized by the channel state and is repeated from
block to block). We will respectively refer to these scenarios as
fast and slow power control (FPC, SPC).

The contributions of this paper are as follows: 1. The framework of
repeated games is applied for the first time to the distributed
energy-efficient power control problem; 2. In the case of FPC, we
derive equilibrium PC strategies which are based on a cooperation
plan between the transmitters, Pareto-efficient, and only require
individual channel state information (CSI) at the transmitters and a
public signal to be implemented; 3. In the case of SPC, which is
much more difficult to treat properly, only the set of possible
equilibrium utilities of the stochastic RG (which can be seen as a
counterpart of a capacity region of a distributed channel when
Shannon transmission rate is considered for the utilities) is
derived by exploiting a very recent result in game theory derived by
H\"{o}rner et al. \cite{HornerSugayaTakahashiVieille09} and
Fudenberg and Yamamoto \cite{FudenbergYamamoto09}. To achieve this
utility region, global CSI and a public signal are assumed at the
transmitters. The determination of the equilibrium strategies is
left as a non-trivial extension of this work.

%A short summary and possible extension of this work are provided in the last section.

%------------------------------------------------------------
%------------------------------------------------------------
%------------------------------------------------------------
%------------------------------------------------------------
\section{Signal model}
\label{sec:system-model}

We consider a distributed MAC with a finite number of users, which
is denoted by $K$. The network is said to be distributed in the
sense that the receiver (e.g., a base station) does not dictate to
the transmitters (e.g., mobile stations) their PC policy. Rather,
all the transmitters chooses their policy by themselves and want to
selfishly maximize their energy-efficiency; in particular they can
ignore some specified centralized policies. We assume that the users
transmit their data over block flat fading channels. The equivalent
baseband signal received by the base station can be written as
\begin{equation}
\label{eq:received-signal}
 y= \sum_{i=1}^{K} g_i x_i + z
\end{equation}
with $i  \in \mc{K}$, $\mc{K} = \{1,...,K\}$, $\mathbb{E}|x_i|^2 =
p_i$, $z \sim \mathbb{C}\mathcal{N}(0, \sigma^2)$. Each channel gain
$g_i$ varies over time following a Markov chain and is assumed to be
constant over each block. For each transmitter $i$, the channel gain
$g_i$ is assumed to lie in a discrete set (e.g., because channel
chains are quantized). The notation $\eta_i = \left|g_i \right|^2$,
with $\eta_i \in \Gamma_i, \ \left| \Gamma_i \right| < +\infty$,
will be used. For the transmit power levels $p_i$ they will be
assumed to lie in a compact set $\mc{P}_i = [0, P_i^{\max}]$  in
Sec. \ref{sec:Fast-Power} and in a discrete set $\mc{P}_i =
\left\{P_i^1, ..., P_i^M \right\}$, with $P_i^{M} = P_i^{\max}$, in
Sec. \ref{sec:Slow-Power}. The discrete set assumption is suited to
exploit the results of
\cite{HornerSugayaTakahashiVieille09}\cite{FudenbergYamamoto09}
without introducing additional technicalities we wanted to avoid in
this (relatively short) paper. At last, the receiver is assumed to
implement single-user decoding.

%%%%%%%%%%%%%%%%%%%%%%%%%%%%%%%%%%%%%%%%%%%%%%%%%%%%%%%%%%%
%%%%%%%%%%%%%%%%%%%%%%%%%%%%%%%%%%%%%%%%%%%%%%%%%%%%%%%%%%%
%%%%%%%%%%%%%%%%%%%%%%%%%%%%%%%%%%%%%%%%%%%%%%%%%%%%%%%%%%%
\section{Fast power control and standard repeated games}\label{sec:Fast-Power}

%------------------------------------------------------------
\subsection{Review of the one-shot power conrol game}
\label{sec:game-definition}

Here we review a few key results from \cite{goodman-pc-2000}
concerning the static PC game. We denote by $R_i$ the transmission
information rate (in bps) for user $i$ and $f$ an efficiency
function representing the block success rate, which is assumed to be
sigmoidal and identical for all the users. For a given block, the
signal-to-interference plus noise ratio (SINR) at receiver $i \in
\{1,...,K\}$ is denoted by $\mathrm{SINR}_i$ and writes as:
\begin{equation}
\label{eq:sinr-ne} \mathrm{SINR}_i=\frac{p_i \eta_i}{\sum_{j \in
\mc{K} \backslash i} p_j \eta_j +\sigma^2}
\end{equation}
where $p_i \in [0, P_i^{\max}]$. With these notations, the static PC
game, denoted by $\mc{G}$, is defined in its normal form as follows.

\begin{definition}[Static PC game] \emph{The static PC game is a triplet
$\mc{G} = (\mc{K}, \{\mc{P}_i\}_{i\in\mc{K}},\{u_i\}_{i\in\mc{K}})$
where $\mc{K} = \{1,...,K\}$ is the set of players,
$\mc{P}_1,...,\mc{P}_K$ are the corresponding discrete sets of strategies,
$\mc{P}_i = \{0,\ldots, P_i^{\mathrm{max}}\}$, $P_i^{\mathrm{max}}$ is the
maximum transmit power for player $i$, and $u_1,...,u_k$ are the
utilities of the different players which are defined by:}
\begin{eqnarray}
\label{eq:def-of-utility} u_i(p_1,...,p_K)= \frac{R_i
f(\mathrm{SINR}_i)}{p_i} \ [\mathrm{bit} / \mathrm{J}].
\end{eqnarray}
\end{definition}
We suppose from now that the above description of the game
 is common knowledge and the players are rational (every player does the best for himself
and knows the others do so and so on). An important game solution
concept is the Nash equilibrium (i.e., a point from which no player has interest in
unilaterally deviating). When it exists, the non-saturated
Nash equilibrium  of this game is given by
\begin{equation} \forall i \in \{1,...,K \}, \
p_i^{*}= \frac{\sigma^2}{\eta_i} \frac{\beta^*}{1-(K-1)\beta^{*}}
\label{eq:NE-power}
\end{equation}
where $\beta^{*}$ is the unique solution of the equation
$xf'(x)-f(x)=0$. By using the term ``non-saturated NE'' we mean that
the maximum transmit power for each user, denoted by $
P_i^{\mathrm{max}}$, is assumed to be sufficiently high for not
being reached at the equilibrium i.e., each user maximizes his
energy-efficiency for a value less than $P_i^{\mathrm{max}}$ (see
\cite{lasaulce-twc-2009} for more technical details about this
assumption). An important property of the NE given by
(\ref{eq:NE-power}) is that transmitters only need to know their
individual channel gain (i.e., $g_i$) to play their equilibrium
strategy. One of the interesting results we want to prove is that it
is possible to obtain a more efficient equilibrium point when
transmitters can update their powers several times per block while
keeping this key information property of individual CSI.

% EXTENSION: FIND THE EQUILIBRIUM STRATEGIES. EQUAL SINR? OPERATING POINT?

%----------------------------------------------------
\subsection{The discounted repeated power control game}

In this section, we assume that the transmitters can update
their powers within time intervals less than the channel coherence
time. The instants at which the
transmitters update their powers are called game stages. Therefore,
for each channel realization $\ul{g} = (g_1,...,g_K)$ a given
repeated game is played. As mentioned in Sec. \ref{sec:intro}, the
fact that the PC game is repeated induces new behaviors (namely
cooperative behaviors) for the transmitters. Because of repetition,
selfish but efficient agreements between transmitters are possible.
In this work, we propose an operating point (OP) of the one-shot PC
game which can serve as a part of a cooperation plan between the
transmitters. Before defining the repeated power control game, let
define the proposed OP.

By considering all the points $(p_1,...,p_K)$ such that $p_i \in [0,
P_i^{\mathrm{max}}]$, $i \in \mc{K}$, one obtains the feasible
utility region. We consider a subset of points of this region for
which the power profiles $(p_1,...,p_K)$ verify
$p_i|g_i|^2=p_j|g_j|^2$ for all $(i,j) \in \mc{K}^2$. Such a subset is made of the following system
of equations:
\begin{equation}
\label{eq:def-operating-point} \forall (i,j) \in \mc{K}^2, \
\frac{\partial u_i}{\partial p_i}(\ul{p}) = 0 \;\;\text{with
}p_i|g_i|^2=p_j|g_j|^2.
\end{equation}
It turns out that, following the lines of the proof of SE uniqueness
in \cite{lasaulce-twc-2009}, it is easy to show that a sufficient
condition for ensuring both existence and uniqueness of the solution
to this system of equations is that there exists $ x_0 \in
]0,\frac{1}{K-1}[$ such
 that $\frac{f''(x)}{f'(x)}-\frac{2(K-1)}{1-(K-1)x}$ is strictly positive on $]0,x_0[$ and strictly negative
 on $]x_0,\frac{1}{K-1}[$. It is satisfied for the two efficiency functions the authors are aware of, which are:
$f(x)=(1-e^{-x})^M$ \cite{shah-pimrc-1998} and $f(x) =
e^{-\frac{c}{x}}$ \cite{belmega-valuetools-2009}
 with $c = 2^R -1$ ($R$ is the transmission
rate). Under the aforementioned condition, the unique solution of
(\ref{eq:def-operating-point}) can be checked to be:
\begin{equation}
\label{eq:power-profile-op-pt} \forall i \in \mc{K}, \
p_i^{\mathrm{OP}} =
\frac{\sigma^2}{\eta_i}\frac{\gamma^{*}}{1-(K-1)\gamma^{*}}
\end{equation}
where $\gamma^{*}$ is the unique solution of $ x[1-(K-1)\cdot x]
f'(x)-f(x)=0$. The proposed OP, given by
(\ref{eq:power-profile-op-pt}), is thus fair in the sense of the
SINR since $\forall i \in \mc{K}$, $\mathrm{SINR}_i = \gamma^{*}$.
We are going to exploit this point of the one-shot
PC game to build equilibrium strategies of the DRG.\\
Let us define a strategy for the discounted repeated game. The
transmitters are assumed to receive a public signal $s(t)$ after
playing at game stage $t$ and keep this in memory. This public
signal is linked to the actions of the transmitters by an
observation function $\phi: \mc{P}_1 \times...\times \mc{P}_K
\longrightarrow \mc{S}$.
\begin{definition}[Players' strategies in the RG] \emph{A pure strategy for
player $i \in \mc{K}$ is a sequence of causal functions
$\left(\tau_{i,t} \right)_{t \geq 1}$ with}
\begin{equation}
\label{eq:strategy-rg} \tau_{i,t}: \left|
\begin{array}{ccc}
\mc{H}_t & \rightarrow & [0, P_i^{\mathrm{max}}]\\
 \ul{h}_t & \mapsto & p_i(t)
\end{array}
\right.
\end{equation}
\emph{where $t$ is the game stage index, $\ul{h}_t =
(s(1),...,s(t-1))$ is the game history vector and $\mc{H}_t =
\mc{S}^{t-1}$.}
\end{definition}
The strategy of player $i$, which is a sequence of functions, will
be denoted by $\tau_i$. The vector of strategies $\ul{\tau} =
(\tau_1, ..., \tau_K)$ will be referred to a joint strategy. A joint
strategy $\ul{\tau}$ induces in a natural way a unique action plan
$(\ul{p}(t))_{t\geq 1}$. To each profile of powers $\ul{p}(t)$
corresponds a certain instantaneous utility $u_i(\ul{p}(t)) $ for
player $i$. In our setup, each player does not care about what he
gets at a given stage but what he gets over the whole duration of
the game. This is why we consider a utility function resulting from
averaging over the instantaneous utility.
\begin{definition}[Players' utilities in the RG] \emph{Let $\ul{\tau} = (\tau_1, ..., \tau_K)$
be a joint strategy. The utility for player $i \in \mc{K}$ is
defined by:}
\begin{equation}
 v_i^{\lambda}(\ul{\tau})  = \sum_{t =1}^{\infty}
\lambda (1 - \lambda)^{t-1} u_i(\ul{p}(t))
\end{equation}
\emph{where $\ul{p}(t)$ is the power profile of the action plan
induced by the joint strategy $\ul{\tau}$ and $0<\lambda<1$ is a
parameter of the DRG called the discount factor and is known to
every player (since the game is with complete information).}
\end{definition}
In the current available wireless
  literature on the problem under
investigation discounted repeated games (DRG) are used as follows:
in \cite{etkin-jsac-2007} the discount factor is used as a way of
accounting for the delay sensitivity of the network; in
\cite{wu-twc-2009} the discount factor is used to let the
transmitters the possibility to value short-term and long-term gains
differently. Interestingly, \cite{Sorin92}\cite{Shapley53} offers
another interpretation of this model. Indeed, the author sees the
DRG as a finite RG where the number of game duration would be
unknown to the players and considered as an integer-valued random
variable, finite almost surely, whose law is known by the players.
Otherwise said, $\lambda$ can be seen as the stopping probability at
each game stage: the probability that the game stops at stage $t$ is
thus $\lambda (1-\lambda )^{ t-1 }$. The function $v_i^{\lambda}$
would correspond to an expected utility given the law of the game
duration. This shows that the discount factor is also useful to
study wireless games where a player enters/leaves the game.
\begin{theorem}[Equilibrium strategies in the DRG]\label{theo:eq-strategies} \emph{Assume that
the following condition is met:}
\begin{equation}
\label{eq:cond2} \lambda \leq \frac{1-(K-1)\gamma^{*}}{
 (K-1)\gamma^{*}} \frac{f(\gamma^{*})}{f(\beta^*)}
 -\frac{1-(K-1)\beta^*}{(K-1)\beta^*}.
\end{equation}
\emph{Then, for all $i \in \mc{K}$, the following action plan is a
subgame perfect NE of the DRG for any distribution for the channel
gains:}
\begin{equation}
\forall t \geq 1, \ \tau_{i,t} = \left|
\begin{array}{ll}
p_i^{\mathrm{OP}} & \text{ \emph{if the other players play} } \tilde{p}_{-i}\\
p_i^* &\text{ \emph{otherwise}}
\end{array}.
\right. \label{eq:eq-strategies}
\end{equation}
\end{theorem}
The proof of this theorem is not provided here; the main idea
of the proof is to derive a sufficient condition on the discount
factor such that the maximum gain induced by a unilateral deviation
is less than the loss induced by the punishment procedure that the
other transmitters apply by playing at the one-shot game NE. The
proposed cooperation plan therefore consists in playing at the
operating point if no transmitter deviates from this point. If one
transmitter deviates from the OP, then all the other transmitters
play the action corresponding to one-shot game NE. At this point it
is possible to see very clearly the information assumptions needed
to implement the proposed distributed power control policies. To
play at $p_i^{\mathrm{OP}}$ or $p_i^*$ only the individual CSI
($\eta_i = |g_i|^2$) is needed by each transmitter. To detect the
deviation of a transmitter we propose the following mechanism: the
receiver broadcasts the public signal $s(t) =\sigma^2 + \sum_{i=1}^K
\eta_i(t) p_i(t) \in \SSS$ (note that the knowledge of the individual SINR is a
sufficient condition to re-construct this public signal). At the OP,
this signal equals $\frac{2 \sigma^2}{1 - (K-1) \gamma^{*}}$. Thus,
if one transmitter deviates all the other transmitters detect this
unilateral deviation and can therefore stop cooperating and start
playing the one-shot game NE. Interestingly, the proposed
equilibrium strategies have been found to be Pareto-optimal for all
simulations we have performed. As a result, the proposed PC policies
are both efficient and rely on reasonable information assumptions.
For comparison, the policies based on pricing
\cite{saraydar-com-2002} require global CSI.

% REMARK: PUBLIC SIGNAL IN TERMS OF SIGNALLING: symbol rate, not good.

%%%%%%%%%%%%%%%%%%%%%%%%%%%%%%%%%%%%%%%%%%%%%%%%%%%%%%%%%%%
%%%%%%%%%%%%%%%%%%%%%%%%%%%%%%%%%%%%%%%%%%%%%%%%%%%%%%%%%%%
%%%%%%%%%%%%%%%%%%%%%%%%%%%%%%%%%%%%%%%%%%%%%%%%%%%%%%%%%%%
\section{Slow power control and stochastic
discounted repeated games}\label{sec:Slow-Power}

From now on, we consider a more general scenario in which channel
gains $\ul{\eta}$ can vary from game stage to game stage. The utility function
at a given game stage therefore depends not only on the profile of
actions $\ul{p}(t)$ played at stage $t$ but also on the vector of
channel gains $\ul{g}(t) = (g_1(t), ..., g_K(t))$ and more precisely
on $\ul{\eta}(t) = (\eta_1(t), ..., \eta_K(t)) \in \Gamma$, with
$\Gamma = \Gamma_1 \times ... \times \Gamma_K $. The corresponding
game-theoretic framework is the one of stochastic repeated games.
Our objective is to characterize the set of equilibrium utilities of
the repeated game. This can be thought of as a counterpart of a
capacity region in information theory. The corresponding result is
called a Folk theorem. It turns out that no general Folk theorem is
available for stochastic RG. It is only very recently that some
authors \cite{HornerSugayaTakahashiVieille09}\cite{FudenbergYamamoto09}
 succeeded to derive a Folk theorem for
stochastic RG with public information. To be able to exploit these
very interesting results we assume that every transmitter knows the public
signal $s \in \SSS$ as in the previous section
and has global CSI $\ul{\eta}$ at each game
instance or stage. An important
 condition which is assumed to be satisfied by the channel gain process
  is the irreducibility property.
 \begin{definition}\label{def:irreducibility}
 \emph{Let $\ul{\eta}$ and $\ul{\eta}'$ be two channel states and $\pi(\ul{\eta}'|\ul{\eta})$, the probability that the next state will be $\eta '$ knowing that the actual state is $\eta$. The transition probability $\pi$ is irreducible if for any channel states
  $\ul{\eta}$ and $\ul{\eta}'$, we have $\pi(\ul{\eta}'|\ul{\eta})>0$.}
\end{definition}
The mobility in wireless communication impose that for a given channel
realization, there is always a positive probability for each channel gain to be drawn in the
 next stage. In order to characterize the set of equilibrium payoff, we
  assume that the transition probability is irreducible.
As in the previous section, we assume that the player does not
observe perfectly the actions played by the other player in the past
stages (imperfect monitoring) but have only access to the public
signal $s(t)\in \SSS$.
\subsection{The game course}
\label{sec:RG-game-course}
The game starts at stage $t=1$ with an initial state $g(1)$ which is
known by the players. The transmitters simultaneously choose a power
level $\ul{p}(1)=(p_1(1),\ldots,p_K(1))$ and get a public signal
$s(1)\in \SSS$ from $\phi(\ul{p}(1))$. The stage utility, denoted by
$u_i(\ul{p}(1),\ul{\eta}(1))$ is not known by the player $i$. After
the stage $t-1$, the channel states are drawn from the probability
distribution $\pi(\cdot|\ul{\eta}(t-1)) \in \Delta(\Gamma)$ and the
realization is publicly announced :
$\ul{\eta}(t)=(\eta_1(t),\ldots,\eta_K(t))$. Taking into account the
past history of the game, the players choose simultaneously their
action $p_i(t)$ and get a public signal $s(t)\in \SSS$ from
$\Phi(\ul{p}(t))$ and does not know their stage utility
$u_i(\ul{p}(t),\ul{\eta}(t))$, and so on.
We define the vector of private $\ul{\tilde{h}}_i(t)$ and public
$\ul{\tilde{h}}(t)$ history of player $i$ :
\begin{footnotesize}
\begin{eqnarray*}
\label{eq:private-history-vector} \ul{h}_i(t) &=& (p_i(1),s(1),\ul{\eta}(1),...,p_i(t-1),s(t-1),\ul{\eta}(t-1),\ul{\eta}(t))\\
\label{eq:public-history-vector2} \ul{h}(t) &=& (s(1),\ul{\eta}(1),...,s(t-1),\ul{\eta}(t-1),\ul{\eta}(t))
\end{eqnarray*}
\end{footnotesize}
We define the public history of the game as the intersection of all
private histories.
 Note that the private history contains the public one and the sequence of transmission power $(p_i(t))_{T-1\geq t\geq1}$ of player $i$.
The vector $\ul{h}(t)$ lies in the set
\begin{equation}
\mc{\widetilde{H}}_t  = \left(\mc{S} \times \Gamma
\right)^{t-1}\times \Gamma
\end{equation}
where the notation $(.)^{t-1}$ refer to the Cartesian product of
sets. This  vector (\ref{eq:public-history-vector2}) that is assumed
to be known by each transmitters before playing for block $t$.
The private and public histories are introduced in order to define the private and the public strategies. In the sequel we will restrict ourself only to the public strategies for which it is possible to characterize the set of equilibrium utilities. Note that this restriction does not affect the final result in terms of set of equilibrium utilities. In fact, we show that, in our framework, the players should not take into account their private history. A strategy is a sequence of functions from the history of the game onto a probability distribution over the set of power.
\begin{definition}[Players' strategies in the RG] \emph{A public strategy for
player $i \in \mc{K}$ is a sequence of functions $\left(\tilde{\tau}_{i,t}
\right)_{t \geq 1}$ with}
\begin{equation}
\label{eq:strategy-rg} \widetilde{\tau}_{i,t}: \left|
\begin{array}{ccc}
\mc{\widetilde{H}}_t & \rightarrow & \Delta(\PP_i) \\
 \ul{\tilde{h}}_t & \mapsto & p_i(t).
\end{array}
\right.
\end{equation}
Where $\Delta(\PP_i)$ denote the set of probability over $\PP_i$.
\end{definition}
The public strategy of player $i$ will therefore be denoted by $\tilde{\tau}_i$
while the vector of public strategies $\ul{\tilde{\tau}} = (\tilde{\tau}_1, ..., \tilde{\tau}_K)$
will be referred to a joint public strategy. A joint public strategy
$\ul{\tilde{\tau}}$ induce in a natural way a unique probability $\PPP_{\ul{\tilde{\tau}},\pi}$ over the set of action plans $(\ul{p}(t))_{t\geq
1}$ and sequence of signals $(\ul{s}(t))_{t\geq1}$.
 The averaged utility for player $i$ can
then be defined as follows.

\begin{definition}[Players' utilities in the RG] \emph{Let $\ul{\tilde{\tau}} = (\tilde{\tau}_1,
 ..., \tilde{\tau}_K)$
be a joint mixed strategy. The utility for player $i \in \mc{K}$ if
the initial channel state is $\eta(1)$, is defined by:}
\begin{equation}
\label{eq:utility-rg} \tilde{v}_i(\ul{\tilde{\tau}}, g) = \sum_{t
\leq 1} \lambda (1 - \lambda)^{t-1}
\E_{\ul{\tilde{\tau}},\pi}\left[u_i(\ul{p}(t),
\ul{\eta}(t))|\ul{\eta}(1)\right]
\end{equation}
\emph{where $(\ul{p}(t))_{t \geq 1}$ is the sequence of power
profile induced by the joint strategy $\ul{\tilde{\tau}}$.}
\end{definition}
We present now the proper definition of a stochastic repeated game.
\begin{definition}[Stochastic RG with Public Monitoring]
\emph{A stochastic repeated game with public monitoring is defined
as
$\G=(\mc{K},(\widetilde{\T}_i)_i,(\tilde{v}_i)_i,(\Gamma_i)_i,\pi,\mc{S},
\Phi)$, where $\mc{K}$ is the set of players, $\widetilde{\T}_i$ is
the set of strategy of player $i$, $\tilde{v}_i$, her long-term
utility function, $\pi$ is the transition probability over the set
of channels gains $(\eta_i)_i$, $\Phi$ is the public observation
function and $\mc{S}$ is the set of public signals.}
\end{definition}
We suppose from now that the above description of the game is common knowledge and the
 players are rational (every player does the best for himself
and knows the others do so and so on).

\subsection{Equilibrium concept}
\label{sec:equilibrium-concept}
At this point, public Nash equilibrium strategies of the stochastic repeated game starting with the channel state $g$ can be
defined.
\begin{definition}[Public Equilibrium Strategies of the RG] \emph{A public
mixed strategy $\ul{\tilde{\tau}}$ supports an equilibrium of the
stochastic repeated game with initial channel state $\ul{\eta}(1)$
if}
\begin{equation}
\forall i \in \mc{K}, \forall \tilde{\tau}_i', \
\tilde{v}_i(\ul{\tilde{\tau}},\ul{\eta}(1)) \geq
\tilde{v}_i(\tilde{\tau}_i', \ul{\tilde{\tau}}_{-i},\ul{\eta}(1))
\end{equation}
where $-i$ is the standard notation to refer to the set $\mc{K}
\backslash \{i\}$; here $\ul{\tilde{\tau}}_{-i} = (\tilde{\tau}_1,
..., \tilde{\tau}_{i-1}, \tilde{\tau}_{i+1},...,\tilde{\tau}_K)$.
\end{definition}
The notion of Nash equilibrium in repeated game is refined by the
sub-game perfection property, introduced by Selten for extensive
games \cite{selten65}, \cite{selten75}. For a sub-game perfect equilibrium, the incentives hold along the duration of the game.
\begin{definition} [Perfect Public Equilibrium Strat. of
the RG]\label{sec:RG-equilibrium}
\emph{A public strategy profile $\ul{\tilde{\tau}}$ is a perfect public
 equilibrium if for every $h^t \in \mc{\widetilde{H}}_t $, the continuation
  profile $\ul{\tilde{\tau}}_{|h^t}$ is a Nash equilibrium of the restricted
  stochastic repeated game starting with the channel state $g(t)$. We
  denote $E_{\lambda}(\eta(1))$ the set of Perfect public equilibrium of the game with initial state $\eta(1) \in \Gamma$ and discount factor $\lambda$.}
\end{definition}
An important issue is precisely to characterize the set of possible
 equilibrium payoff or public perfect equilibrium payoff in the repeated game. This kind of result
  often appears as ``Folk Theorem'' (see e.g.,\cite{Aum81}\cite{Sorin92}). A huge part of the
  literature is dedicated to find the set of equilibria under different assumptions, but a
  general characterization is still unavailable. Our model is included in the framework of
   stochastic repeated game with imperfect public monitoring.
\subsection{Independence of the initial State}
In classical models of stochastic repeated game, the
 initial state $\eta(1)$ could be determinant for
 characterizing the solutions of our problem. However, it is
  natural to think that the initial state of channel
  gain will not influence the future sequence of channel
  realization. We present some results of Dutta (1995)
   \cite{Dutta95} that formalize the above statement. Because
    of the irreducibility property of the channel stochastic
     process, the limit set of feasible utilities, the
      set of perfect public equilibrium utilities and
      the minmax utilities are independent of the initial state.
\begin{theorem}[Independence of the Initial State]
Suppose that the stochastic repeated game is irreducible (\ref{def:irreducibility}), it implies that :
\emph{\begin{itemize}
\item The limit of the minmax is independent of the initial state i.e. $\lim_{\lambda \longrightarrow 0}min_{\tau_{-i}}max_{\tau_{i}} \tilde{v}_i(\tilde{\tau}_i,\ul{\tilde{\tau}}_{-i},\eta(1))=\tilde{v}_i$ for all $\ul{\eta}(1)$ and all $i \in K$.
\item The limit set of feasible utilities is independent of the initial state i.e. $\lim_{\lambda \longrightarrow 0} F_{\lambda}(\ul{\eta}(1))=F$ for all $\eta(1)$.
\item The limit set of public perfect equilibrium utilities is independent of the initial state i.e. $\lim_{\lambda \longrightarrow 0} E_{\lambda}(\eta(1))=E$ for all $\eta(1)$.
\end{itemize}}
\end{theorem}
The following definition is fundamental for characterize
 the set of public perfect equilibrium payoff of our repeated game.
\begin{definition}
\emph{We define the set of asymptotic feasible and individually
rational payoff by}: \begin{equation}F^*=\{x \in F| x_i\geq
\tilde{v}_i,\; \forall i \in K\}\end{equation}
\end{definition}
The set $F^*$ is defined as the set of energy-efficiency utilities the
players can get such that each of them has more than his minmax utility.

\subsection{Main Result : Folk Theorem}
\label{sec:main-result}

The following theorem state that only a condition over
 the discount factor $\lambda$ is sufficient to have a
  sub-game perfect equilibrium property for a utility vector $\ul{u}$ in $F^*$.
\begin{theorem}
\emph{For each utility vector $\ul{u} \in F^*$, there exists
 a $\lambda_0$ such that for all $\lambda < \lambda_0$, there
  exists is a perfect public equilibrium strategy of our
  stochastic repeated power control game, such that the long-term utility equals $\ul{u}\in F^*$.}
\end{theorem}
The proof is based on H\"{o}rner, Sugaya, Takahashi and
Vieille (2009), \cite{HornerSugayaTakahashiVieille09}
%; Fudenberg and Yamamoto (2009) \cite{FudenbergYamamoto09}
 ; Kandori and Matsushima (1998) \cite{KandoriMatsushima98}.

\section{Numerical Illustration of Optimal Equilibrium Utilities}
\label{sec:numerical-results}

\begin{figure}%[ht]
\centering
\includegraphics[width=0.43\textwidth]{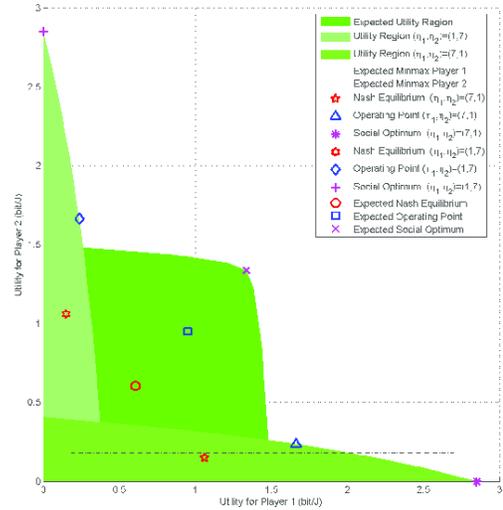}
%\vspace{-0.8cm}
\caption{Utility regions for $(K,M,N)=(2,2,2)$ considering different channels configurations and their expected utilities. Our procedure lead to the expected social optimum.}
\label{fig:utility-region-222}
\end{figure}

\begin{figure}%[ht]
\centering
\includegraphics[width=0.4\textwidth]{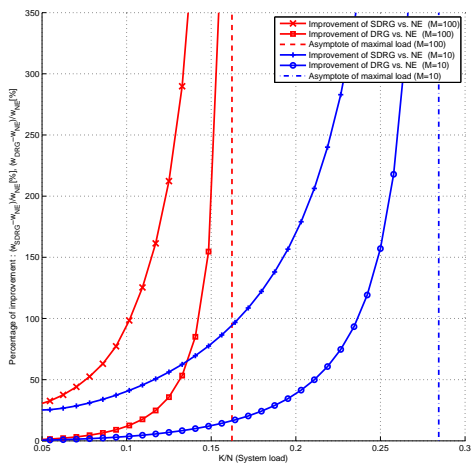}
%\vspace{-0.8cm}
\caption{The gain brought by the repeated-game based cooperation in
terms of the sum of utilities (standard and stochastic repeated
games) w.r.t. to the purely non-cooperative scenario (Nash
Equilibrium).} \label{fig:efficiency-sw}
\end{figure}

The above result implies that each Pareto-optimal utility vector that is
 individually rational can be sustained by a public perfect equilibrium
  strategy for a discount factor sufficiently small. In practice, we
   have to focus on a particular Pareto-optimal point which
    is individually rational. For example, denote by $\tilde{p}$ the solution of the maximization problem : $\max_{p \in P \text{ s.t. } u_i\geq v_i \forall i \in K}\sum_{i \in K} \alpha_i u_i(p)$ and $\tilde{u}$ it's corresponding utility vector. The above theorem states that $\tilde{u}$ is a public perfect equilibrium utility of the $\lambda-$discounted repeated game for a sufficiently small discount factor if $\tilde{u}$ Pareto-dominates the Minmax utilities.  In the whole section we consider the same type of scenarios as
\cite{meshkati-jsac-2006}\cite{lasaulce-twc-2009} namely random code
division multiple access systems with a spreading factor equal to
$N$ and the efficiency
function $f(x) = (1- e^{-x})^M$, $M$ being the block length. \\
$\square$ We consider a simple stochastic process with two channel
states: $(\eta_1,\eta_2)\in \{(7,1),(1,7)\}$. The transition
probability is constant over the channel states: $\pi
(\cdot)=(\frac{1}{2},\frac{1}{2})$ and its
 invariant measure is
$\mu=(\frac{1}{2},\frac{1}{2})$. Consider the scenario $(K,M,N) =
(2,2,2)$.  Fig. \ref{fig:utility-region-222} represents the
achievable utility region for the two different channel state and
the long-term expected utility region. The positive orthan denotes
the set of expected individually rational utilities. Its
intersection with the expected achievable utility region describes
the set of public perfect equilibrium utility. Three important
points are highlighted in the different scenario: the expected Nash
equilibrium of the one-shot game studied in \cite{goodman-pc-2000},
the expected operating/cooperation point studied by Le Treust and
Lasaulce 2010 \cite{LeTreustLasaulce(PowerControlRG)10}, and the
point where the expected social welfare (sum of utilities) is
maximized (star). From this figure it can be seen that: a
significant gain can be obtained by using a model of repeated
games instead of the one-shot model. Moreover, significant improvement in term of expected utilities is a direct consequence of the full CSI instead of individual CSI.\\
$\square$ As a second type of numerical results, the performance gain
brought by the stochastic discounted repeated game (SDRG) formulation of the distributed PC problem is
assessed. Considering a simple stochastic process where $\eta_i=2$ and $\eta_{j}=1$ for all $j \in \mathcal{K}  \backslash\{i\}$ and the $i$'s player is drawn with uniform distribution over the $K$ players. We compute the expected social utility the players get at the social optimum $w_{SDRG}$. Denote by $w_{NE}$ (resp. $w_{DRG}$ and $w_{SDRG}$) the
efficiency of the NE (resp. DRG and SDRG equilibrium) in terms of
social welfare i.e. the sum of utilities of the players. Fig.
\ref{fig:efficiency-sw} represents the quantity
$\frac{w_{SDRG}-w_{NE}}{w_{NE}}$ and $\frac{w_{DRG}-w_{NE}}{w_{NE}} $
in percentage as a function of the spectral efficiency $\alpha =
\frac{K}{N}$ with $N=128$ and $2\leq K<\frac{N}{\beta^*}+1$. The
asymptotes $\alpha_{max} = \frac{1}{\beta^*}+\frac{1}{N}$ are
indicated by dotted lines for different values $M \in
\{10,100\}$. The improvement becomes very significant when the system load
is close to $\frac{1}{N}+\frac{1}{\beta^*}$, this is because the
power at the
 one-shot game NE becomes large when the system becomes more and more
loaded. As explained in \cite{lasaulce-twc-2009} for the Stackelberg
approach these gains are in fact limited by the maximum transmit
power.\\

\section{Conclusion}
\label{sec:conclusion}

Repeating a power control game is a way of introducing cooperation
between selfish transmitters. In this paper, we have shown that the
corresponding cooperative power control policies can be implemented
without using explicit cooperation channels between the
transmitters. In the case of fast PC, only individual CSI
and a realistic public signal are required to implement the proposed
schemes. In the case of slow power control, only the feasible
utility region has been derived, the equilibrium power control
strategies to achieve the corresponding points still need to be
found. Both in the cases of fast and slow power control, the
cooperation gain induced by the underlying cooperation plans is
shown to be significant. The repeated game formulation of the
distributed power control problem therefore shows a way of reaching
interesting trade-offs in terms of global network energy-efficiency
and signalling.

%\bibliographystyle{IEEEtran}
%\bibliography{BiblioMael}

%\begin{figure}%[h]
%\centering
%\includegraphics[width=0.50\textwidth]{BattleSexeEntropy0,5-1,7BW2011-11-14.eps}
%\caption{Achievable utility region for the battle of sexes under capacity constraints.}
%\label{fig:BattleofSexes}
%\end{figure}

%\vspace{-0.9cm}
%\vspace{-0.5cm}

\end{document}